\documentclass[showpacs,preprintnumbers,amsmath,amssymb]{revtex4}

\usepackage{graphicx}
\usepackage{dcolumn}
\usepackage{bm}
\usepackage{epsfig}
\usepackage{subeqn}
\usepackage{epsfig}
\usepackage{amsmath}
\usepackage{amsfonts}

\newcommand{\ba}{\begin{eqnarray}}
\newcommand{\ea}{\end{eqnarray}}
\newcommand{\be}{\begin{equation}}
\newcommand{\ee}{\end{equation}}
\newcommand{\bea}{\begin{eqnarray}}
\newcommand{\eea}{\end{eqnarray}}

\usepackage{theorem}

\theorembodyfont{\rmfamily}
\newtheorem{remark}{Remark}
\theoremstyle{break}

\def\QED{~\rule[-1pt]{5pt}{5pt}\par\medskip}

\def\Id{\mbox{$\bf 1\ $}}

\def\k {\mathfrak{k}}

\begin{document}

\baselineskip .7cm

\title{Switched Control of Electron Nuclear Spin Systems}

\author{Navin Khaneja}

\email{navin@hrl.harvard.edu}

\affiliation{Division of Engineering and Applied Science, Harvard
University, 33 Oxford Street, Cambridge MA 02138}
\date{\today}

\begin{abstract}
In this article, we study control of electron-nuclear spin dynamics 
at magnetic field strengths where the Larmor frequency of the nucleus is 
comparable to the hyperfine coupling strength. The quantization axis for the 
nuclear spin differs from the static $B_0$ field direction and depends 
on the state of the electron spin. The quantization axis can be switched by flipping 
the state of electron spin, allowing for universal control on nuclear spin states.
We show that by performing a sequence of flips 
(each followed by a suitable delay), we can perform any desired rotation on the 
nuclear spins, which can also be conditioned on the state of the electron spin. These 
operations, combined with electron spin rotations can be used to synthesize any unitary 
transformation on the coupled electron-nuclear spin system. 
We discuss how these methods can be used for design of experiments for 
transfer of polarization from the electron to the nuclear spins.
\end{abstract}


\maketitle

\section{Introduction}

In this paper, we study the problem of control of coupled electron-nuclear spin system  
consisting of an electron, coupled with one or more nuclear spins. Manipulation of such 
electron-nuclear spin systems is fundamental in the 
field of ESR/EPR \cite{SJ:2001} with application to the study of structure and dynamics
of paramagnetic species. Coupled electron-nuclear 
spin systems have recently been studied from the perspective of quantum information 
processing \cite{MMS:2003,MSW:2004,Men:2005,MM:2006,HMM:2006b,HMM:2006,Hei:2006, JGP:2004,WJ:2006,GDP:2006,JW:2006, CGDT:2006, Rah:2006,MHS:2006,AF:2007}. Study of control of these systems 
is interesting from perspective of quantum control as dynamics of coupled 
electron-nuclear system has some salient differences compared to coupled nuclear spin 
systems \cite{EBW:1997}. 

The Rabi frequency of the electron (at typically available microwave power in pulsed EPR experiments) is
of orders of magnitude larger than the Rabi frequency of the nucleus 
(at typical rf-power). Duration of $\frac{\pi}{2}$ pulses on the electron and nucleus are in $ns$ and $\mu s$ regimes respectively in EPR experiments. Local rotations on electrons, therefore take much less time than the nucleus. The energy 
eigenstates of the nucleus (at field strengths where hyperfine interactions
and Larmor frequency of the spins is comparable)
depend on the state of the electron and are not aligned with the static $B_0$ 
field. This results in the well studied phenomenon of  
Electron Spin Echo Envelop Modulation (ESEEM) \cite{SJ:2001}.
This opens the possibility of 
controlling the nucleus by simply manipulating the electron spin between its two 
eigenstates (by a series of $\pi$ pulses and delays). The quantization axis and the precession frequency (around this axis) of the nucleus is 
switched every time the 
state of electron spin is flipped, making the nucleus nutate around the 
new quantization axis. We will shown that by performing a sequence of 
flips on the electron (each followed by a suitable delay), we can 
perform any desired rotation on the nuclear spin states. This rotation can also 
be conditioned on the state of the electron spin. This mode of control on nuclear 
spins obtained by switching between two
rotation axis is an excellent example of so called switched control systems in 
control theory \cite{liberzon:03}. Controlling nuclei by switching 
quantization axis is preferable than 
rotating nuclei with rf-fields as hyperfine couplings are orders of 
magnitude larger 
that Rabi frequency obtainable with typical radio frequency power. Therefore much shorter experiments can be designed, which can significantly reduce relaxation losses, mainly arising from short transverse relaxation times of electron. Furthermore, since the 
only manipulation required involves flipping the state of the electron spin, we show that 
it is easy to design these experiments so that they are robust to the Larmor 
dispersion of the electrons. It is expected that this switched mode control of nuclear spins 
will find applications in quantum information processing and pulsed EPR, including
pulsed dynamic nuclear polarization (DNP) experiments \cite{SJ:2001, griffindnp}. 
   
\section{Theory}

We consider as our model, a spin system consisting of one electron spin $S = \frac{1}{2}$ 
and one nuclear spin $I=\frac{1}{2}$ (see, e.g., Sect.~3.5 of \cite{SJ:2001}) in a static magnetic field $B_0$ along the $z$ direction. We will later 
also discuss the case of one electron coupled to many nuclear spins. The Hamiltonian 
$H_0$ of the electron-nuclear spin system in the laboratory frame may be written as 

$$ H_0 = \omega_s S_z + \omega_I I_z + {\bf S \cdot A \cdot I}, $$ 
where $\omega_s$ represents the Larmor precession frequency of the electron spin, 
$\omega_I$ represents Larmor precession frequency of the nuclear spin, and ${\bf A}$ is 
the electron-nuclear hyperfine coupling tensor. Given the Pauli matrices 
$\sigma_{x}:=
\left(\begin{smallmatrix}
0 & 1 \\
1 & 0
\end{smallmatrix}
\right)$,
$
\sigma_{y}:=
i\left(
\begin{smallmatrix}
0 & -1 \\
1 & 0
\end{smallmatrix}
\right)
$
, and
$
\sigma_{z}:=
\left(
\begin{smallmatrix}
1 & 0 \\
0 & -1
\end{smallmatrix}
\right)
$
 and the identity  matrix
$
\sigma_{0}:=
\left(
\begin{smallmatrix}
1 & 0 \\
0 & 1
\end{smallmatrix}
\right)
$,
the operators $S_{j}$ and $I_{k}$ ($j,k \in \{x,y,z\}$) are 
defined by $S_{j}=(\sigma_{j}\otimes\sigma_{0})/2$ and 
$I_{k}= (\sigma_{0}\otimes\sigma_{k})/2$ (see \cite{EBW:1997}).
The static field $B_0$, sets the quantization axis of electron 
spin and the coupling Hamiltonian 
$H_c =  \bf S \cdot A \cdot I $, averages to

$$ H_c = A S_zI_z + B_x S_zI_x + B_y S_z I_y , $$

The transverse plane axis for the nuclear spin subspace 
can be so chosen that the last two terms are combined to form

$$ H_c = A S_z I_z + B S_z I_x, $$

where $B=\sqrt{B_x^2 + B_y^2}$. The full Hamiltonian of the system then takes the form

$$ H_0 = \omega_s S_z + \omega_I I_z + A S_z I_z + B S_z I_x. $$ 

Here $\omega_s = -\gamma_e B_0$ and $\omega_I = - \gamma_n B_0$.
Here $\gamma_e$ is the gyromagnetic ratio of the electron 
(negative) and  $\gamma_n$ is the gyromagnetic ratio 
of the nucleus (we take positive, as for a proton).

Let $\alpha$ and $\beta$ denote the state of the 
electron spin oriented along and opposite to the direction 
of the static  magnetic field $B_0$ respectively. For an 
electron, the $\alpha$ configuration has higher energy than 
the $\beta$ configuration.

When the electron is in the $\alpha$ state, the nucleus sees a 
net field

$$ B_{\alpha} = (B_0 - \frac{A}{2 \gamma_n}) \hat{z} - \frac{B}{2 \gamma_n} \hat{x}.  $$  

\begin{figure}[h]
\begin{center}
\includegraphics[scale=.7]{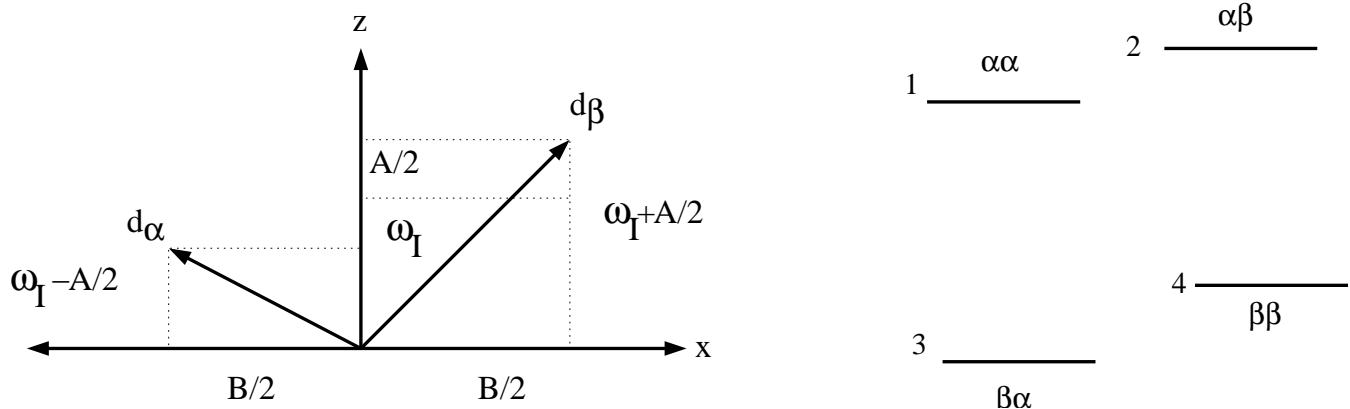}
\end{center}
\caption{Fig. A depicts the vectors $d_{\alpha}$ and $d_{\beta}$ 
when $|\omega_I| > \frac{A}{2}$. Fig. B depicts the corresponding
energy level diagram.}
\label{fig:energy1}
\end{figure}

When the electron is in the $\beta$ state, the nucleus sees a 
net field 
$$ B_{\beta} = (B_0 + \frac{A}{2 \gamma_n}) \hat{z} + \frac{B}{2 \gamma_n}  \hat{x}  $$
The nuclear precession frequency in the two states is given by
\begin{equation}
\label{eq:frequency}
\omega_{\alpha}= \sqrt{(\frac{A}{2} + \omega_I)^2 + \frac{B^2}{4}};\ \omega_{\beta} = \sqrt{(\frac{A}{2}- \omega_I)^2 + \frac{B^2}{4}}.
\end{equation}
where $\omega_I$ is negative. 
Let $d_\alpha$ and $d_\beta$ denote unit vectors along the field directions
$B_\alpha$ and $B_\beta$. The states $|\alpha \alpha \rangle $ and 
$|\alpha \beta \rangle$ represent the states of nuclei oriented along or against 
$d_{\alpha}$ respectively, when the electron is in the $\alpha$ state.
These states constitute the $\alpha$ manifold.
Similarly, the states $|\beta \alpha \rangle $ and $|\beta \beta \rangle $ 
represent the states of nuclei oriented along or against the field 
$d_{\beta}$ respectively, when the electron is in the $\beta$ state.
These states constitute the $\beta$ manifold. 
Fig. \ref{fig:energy1} depicts the field $d_{\alpha}$ 
and $d_{\beta}$ and the energy eigenstates states when 
$|\omega_I| > \frac{A}{2}$. 
 
In the rotating frame (rotating with electron at frequency $\omega_s$), the 
Hamiltonian of the electron-nuclear spin system takes the form

$$ H_{1} = \Omega_S S_z + \omega_I I_z + A S_z I_z + B S_z I_x $$

where $\Omega_S$ is the resonance offset for the electron. 
We assume for now that precession frequency of the electron is well defined 
and $\Omega_S = 0$ (the case when $\Omega_S \neq 0$ will be discussed subsequently).
We now show how by switching between the $\alpha$ and $\beta$ states of the 
electron, by $\pi$ pulses, 
we can synthesize any desired rotation on the spin states of the nuclei.
Furthermore, since $\omega_\alpha \neq \omega_\beta$, we will show
these rotations can be made conditioned on the electronic state.
To fix ideas, consider the state transformation 

$$ |\beta \alpha \rangle \rightarrow  |\beta \beta \rangle . $$

\begin{figure}[h]
\begin{center}
\includegraphics[scale=.5]{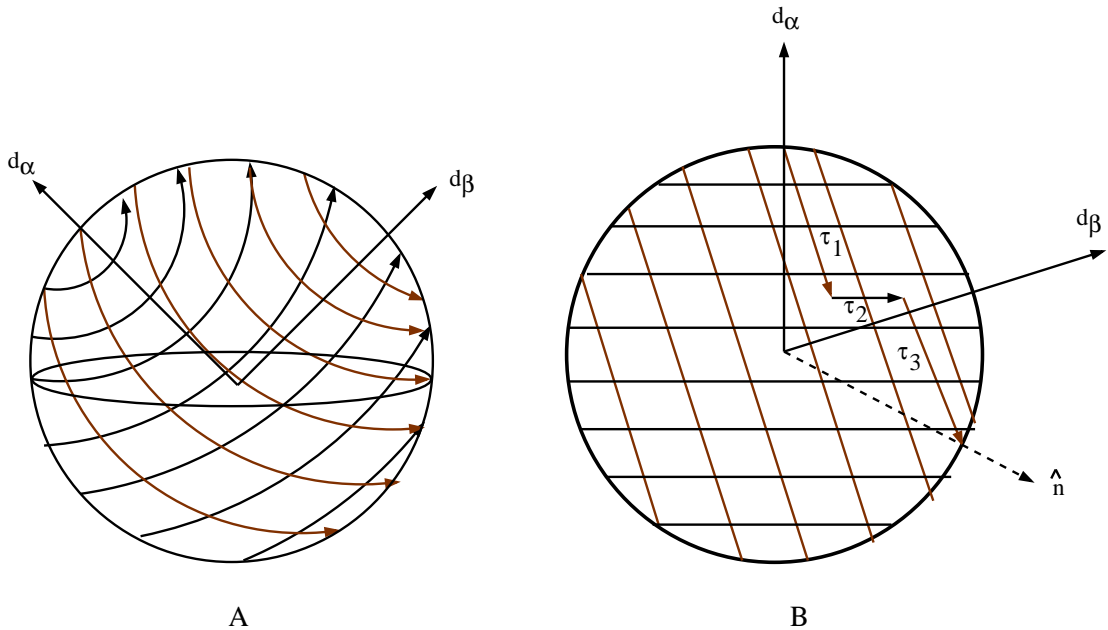}
\end{center}
\caption{ Fig. \ref{fig:traj1}A shows the various trajectories of nuclear spin 
on the Bloch sphere as it precesses around $d_{\alpha}$ and 
$d_{\beta}$ directions. This trajectory plot aids in visual construction of 
rotation between points on the sphere by alternation between rotations 
around $d_{\alpha}$ and $d_{\beta}$.
Backtracking these trajectories can 
help to construct desired rotation of nuclear spin state
by switching between field directions $d_{\alpha}$ and $d_{\beta}$. Fig. 
\ref{fig:traj1}B depicts one such construction (projected on the plane of axis
$d_\alpha$ and $d_\beta$ ) that rotates unit vector 
$d_\alpha$ to the unit direction $\hat n$. The initial rotation is around axis 
$d_\beta$ for time $\tau_1$, followed by rotation around $d_\alpha$ for 
time $\tau_2$ and finally around $d_\beta$ for time $\tau_3$. 
The times $\tau_1$, $\tau_2$
and $\tau_3$ can be explicitly computed from this figure.}
\label{fig:traj1}
\end{figure}

When the electron is flipped from $\beta$ to $\alpha$ state, the effective field
felt by the nucleus switches from direction $d_{\beta}$ to 
$d_{\alpha}$ and the nuclei begin to precess around this new axis.
Following this precession for time $\tau_1$, we flip the electron back, 
causing the precession axis to return to $d_{\beta}$ and then let the 
precession happen for time $\tau_2$ and so on. Since $d_{\alpha}$
and $d_{\beta}$ are two independent axis of rotations, switching 
between them followed by precession can generate any three 
dimension rotation of the nuclear orientation. Fig. \ref{fig:traj1}A shows the 
trajectories of nuclear spin 
on the Bloch sphere as it precesses around $d_{\alpha}$ and 
$d_{\beta}$ directions. These trajectory plots can be used to 
construct the sequence of switchings between the axis (and delays between switchings)
$d_{\alpha}$ and $d_{\beta}$ to produce the 
desired rotation of nuclear spin state. Fig.  \ref{fig:traj1}B, shows
one such graphical construction for finding the switching times 
$\tau_1$, $\tau_2$ and $\tau_3$, that rotates the unit vector $d_\alpha$ to 
vector $\hat n$, by alternate rotations around $d_\beta$ and $d_\alpha$.

\begin{figure}[h]
\begin{center}
\includegraphics[scale=.7]{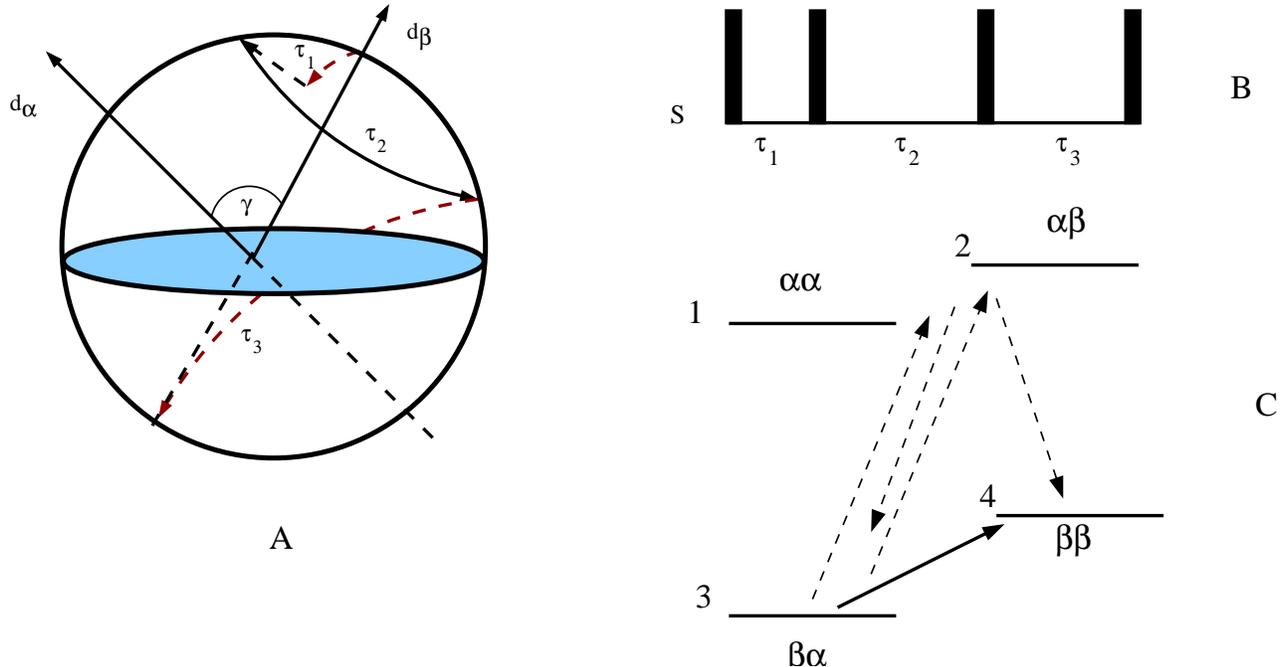}
\end{center}
\caption{The Fig. A depicts the trajectory executed by nuclear spin 
magnetization, initially oriented along the direction $d^{\beta}$, when 
the pulse sequence 
in Fig. B is implemented. The transformation is the inversion of the 
nuclear spin resulting in the transformation 
$|\beta \alpha \rangle \rightarrow  |\beta \beta \rangle$. Fig. C, shows 
the desired rotation (shown with solid arrow) 
$|\beta \alpha \rangle \rightarrow  |\beta \beta \rangle$ 
is implemented by switching the electron between $\alpha$ and $\beta$ 
manifold.}
\label{fig:traj2}
\end{figure}

 Fig. \ref{fig:traj2} shows the transformation 
$|\beta \alpha \rangle \rightarrow  |\beta \beta \rangle,$ with a three 
period pulse sequence as shown in the Fig. \ref{fig:traj2}.
The delays $\tau_1$, $\tau_2$ and $\tau_3$ are chosen so that the nuclear
spin magnetization execute the trajectory shown in Fig. \ref{fig:traj2}A. The 
last $\pi$ pulse in the figure returns the electron to state $\beta$. The times
$\tau_1$, $\tau_2$ and $\tau_3$ can be readily computed from this figure.

Similar arguments can be 
used to transform any energy 
eigenstate into another by simply flipping the electron spins followed by 
appropriate delays. This then naturally leads us to the question of
number of times the state of the electron needs to be switched to 
produce any desired rotation on the nuclear spins, in a given eigenstate 
in either $\alpha$ or $\beta$ manifold. The problem of finding the 
minimum number of 
switchings for producing arbitrary rotations by switching between two axis 
of rotation has been studied before. We recapitulate the main results and 
motivate further extentions.

\begin{remark} {\bf Uniform Generation:} {\rm Let $\gamma$ denote the angle between two rotation 
axis $d_\alpha$ and $d_\beta$ as shown in Fig.(\ref{fig:traj2}). Then it is 
a known result \cite{LOW:1971} that any three dimensional 
rotation in group $SO(3)$ (similar result holds for group $SU(2)$) can be 
constructed by not more that $k+2$ rotations (performed around $d_\alpha$
or $d_\beta$), where $$ \frac{\pi}{k+1} \leq \gamma < \frac{\pi}{k}. $$  
Therefore, starting in either $\alpha$ or $\beta$ manifold , 
any rotation on nuclear spin can be obtained 
by a pulse sequence as shown in  Fig.(\ref{fig:traj2}B) with no more that
$k+2$ evolution periods $\tau_i$. Furthermore, there exist algorithms to find 
the sequence with minimum number of switches \cite{alessandro} 
to generate any arbitrary rotation using only rotations around 
$d_\alpha$ and $d_\beta$. Here, we present a strategy for obtaining such 
a sequence (which need not be minimal in terms of number of switches) which is 
easy to construct geometrically.
 
Let $\tau_1-\tau_k$ ($k$ even) be the times for alternate rotations 
(starting with $d_\beta$) around $d_\beta$ and $d_\alpha$, with angular frequency 
$\omega_\beta$ and $\omega_\alpha$, so chosen that the resulting rotation takes the 
initial vector $d_\alpha$ to unit direction $\hat n$ as in 
figure (\ref{fig:traj1}B). Let $U(\alpha, \tau) = \exp(-i \omega_\alpha \tau\ d_\alpha \cdot \sigma)$ and $U(\beta, \tau) = \exp(-i \omega_\beta \tau\ d_\beta \cdot \sigma)$, be rotations in 
$SU(2)$ around axis $d_\alpha$ and $d_\beta$. A rotation by angle $\theta$ around
an unit vector $\hat n = n_x i + n_y j + n_z k $, takes the form
\begin{equation}
\label{eq:su2}
\exp(-i \theta \hat n \ \cdot \sigma ) = \cos \frac{\theta}{2} \Id -i \ 2 \sin \frac{\theta}{2} \hat n \cdot \sigma, 
\end{equation}where $ \hat n \cdot \sigma = n_x \sigma_x + n_y \sigma_y + n_z \sigma_z $. 
Then, a sequence of rotations
\begin{equation}
\label{eq:rot.tracing}
\underbrace{U(\alpha, \tau_k) \dots U(\beta, \tau_1)}_{U_1} U(\alpha, \tau) \underbrace{U(\beta, -\tau_1) \dots U(\alpha, -\tau_k)}_{U_1^{-1}},
\end{equation} will generate a rotation around the unit vector
$\hat n$ with an angle $\theta = \tau \omega_\alpha$. Let 
\begin{equation}
\label{eq:def1}
t_{\alpha} = \frac{2 \pi}{\omega_{\alpha}};\ \ t_{\beta} = \frac{2 \pi}{\omega_{\beta}};   
\end{equation}

The rotations $U(\beta, -\tau)$ and $U(\alpha, -\tau)$ are simply 
obtained by observing that
\begin{equation}
\label{eq:negation}
U(\beta, t_\beta - \tau) =  -U(\beta, -\tau);\ \ 
U(\alpha, t_\alpha - \tau) =  -U(\alpha, -\tau)
\end{equation} Since $k$ is even, the negative signs in (\ref{eq:rot.tracing}) cancel 
out (if $k$ is odd, it just produces a global phase).
We now show how all these results need to be further generalized if one is 
interested in producing controlled operations on the 
nuclei.}
\end{remark}

\subsection{Controlled Operations on Nuclear Spins }

A $\alpha$ manifold selective controlled operation on the nuclear spins 
transforms $|\alpha \alpha \rangle \rightarrow |\alpha \alpha' \rangle $
and $|\alpha \beta \rangle \rightarrow |\alpha \beta' \rangle$.
While the $\beta$ electron manifold is left unperturbed, i.e., 
$|\beta \alpha \rangle \rightarrow |\beta \alpha \rangle $
and $|\beta \beta \rangle \rightarrow |\beta \beta \rangle $. 
We show that such controlled operations can again be 
performed by toggling the state of the electron spin with $\pi$ pulses,
followed by suitable delays. This is best seen by observing that
by flipping the state of the electron spins, we can switch between the 
effective Hamiltonians (generators)

\begin{equation}
H_{1} = \omega_I I_z + S_z (AI_z + BI_x) 
\end{equation}

and

\begin{equation}
H_{2} = \omega_I I_z - S_z (A I_z + B I_x)
\end{equation}

For example, the pulse sequence $\tau_1-\pi-\tau_2-\pi$ (where $\pi$ pulse is on the electron) evolves $H_1$ for time $\tau_1$ 
followed by $H_2$ for time $\tau_2$. A straightforward calculation shows that when $\omega_\alpha \neq \omega_\beta$ (in Eq. \ref{eq:frequency}),  the repeated commutators of $-iH_1$ and $-iH_2$ generate a six dimensional algebra, spanned by generators $ \k = -i\{I_x, I_y, I_z, S_zI_z, S_zI_x, S_zI_y \}$. Then a standard result on controllability \cite{jurd} states that any unitary transformation generated by 
these generators can be produced by simply switching between $-iH_1$ and $-iH_2$, i.e., 
any rotation of the kind 
\begin{equation}
\label{eq:arbitrary}
\exp(-i (a S_zI_p + b I_q)), 
\end{equation}
where $I_p$ and $I_q$ are nuclear spin operators can be synthesized. Specifically,
we can perform a rotation $\exp(-i S^{\alpha}I_{\gamma})$ that only rotates 
nuclei in the electron's $\alpha$ manifold 
($ S^{\alpha} = (\frac{1}{2} + S_z) $, $S^{\beta} = (\frac{1}{2} - S_z) $ ).

We rewrite, 

$$ H_1 = \omega_I (S_z^{\alpha} + S_z^{\beta})I_z + \frac{(S_z^{\alpha} - S_z^{\beta})}{2}( AI_z + BI_x), $$which when rewritten looks like

$$ H_1 = S_z^{\alpha} (\omega_I I_z + \frac{(AI_z + BI_x)}{2}) + S_z^{\beta}(\omega_I I_z -\frac{( AI_z + BI_x)}{2}), $$

which written in a $6 \times 6$ matrix form takes the block diagonal form

$$ 
H_1 = \left(\begin{array}{cc}
D_{\alpha} & 0 \\
0 & D_{\beta}
\end{array}
\right); \ \  H_2 = \left(\begin{array}{cc}
D_{\beta} & 0 \\
0 & D_{\alpha}
\end{array}
\right), $$ where $D_\alpha$ denotes rotation around axis $d_\alpha$ with angular frequency
$\omega_{\alpha}$ and $D_\beta$ denotes rotation around axis $d_\beta$ with angular frequency
$\omega_{\beta}$. The algebra $\k$ is simply all block diagonal $6 \times 6$ skew 
Hermitian matrices, i.e., matrices of the form $\left(\begin{array}{cc} A_1 & 0 \\ 0 & A_2
\end{array} \right )$, i.e. $A_1 = -A_1^\dagger$ and $A_2 = -A_2^\dagger$. Note, in the special case 
when  $A = 0$ in (\ref{eq:frequency}), then $\omega_\alpha = \omega_\beta$ in (\ref{eq:frequency}). In 
this special case the algebra generated by $H_1$ and $H_2$ is only $3$ dimesional algebra and 
we cannot produce arbitrary controlled operations on the nuclei. This case however is of only 
theoretical interest and we will not consider it any further.

A $\pi$ pulse on the electron will switch $H_1$ to $H_2$ and vice-versa.
A sequence of delays and flips $\tau_1-\pi-\tau_2-\dots-\tau_k-\pi$, where $k$ is say even, 
will execute the following rotations on the nuclear spins in $\alpha$ and $\beta$ manifold 
respectively 

$$ U_\alpha = U(\beta, \tau_k) \dots U(\alpha, \tau_1) ; \ \ U_\beta = U(\alpha, \tau_k) \dots U(\beta, \tau_1). $$Here $U(\alpha, \tau) =  \exp(-i D_\alpha \tau)$ and $U(\beta, \tau) =  \exp(-i D_\beta \tau)$. 
The goal is to find the sequence of times $\tau_k$, such that 
$U_{\beta} = I $  and $U_{\alpha}=U_f$ is the 
desired rotation. This constitutes a controlled rotation on nuclear spins, 
conditioned on the electron state. Similarly, finding a sequence of 
times $\tau_k$, such that  $U_{\beta} = U_{\alpha}=U_f$ constitutes 
a uniform rotation on the nuclei.  In general, we want to find a sequence 
that performs manifold selective operations, i.e., $U_{\alpha}$  and
$U_{\beta}$ are as desired. It is also a known result 
\cite{LOW:1971} that there exists a finite number of switchings
$N$, such that arbitrary $U_{\alpha}$ and $U_{\beta}$ can be constructed by no more than
$N$ number of switchings. Finding methods to generate these in minimum number of switchings 
is an interesting problem. We present a constructive search strategy for 
achieving this (our methods need not be minimal in terms of number of switching).

Let us perform a series of rotations on the $\beta$ manifold (with $k$ even) given by 
$$ U_{\beta} = \underbrace{U(\beta, -\tau_1) \dots U(\alpha, -\tau_k)}_{U_1^{-1}} \underbrace{U(\alpha, \tau_k) \dots U(\beta, \tau_1)}_{U_1}, $$ where $U_{\beta}$ is implemented as
 $$ U_{\beta} = U(\beta, t_\beta-\tau_1) \dots U(\alpha, t_\alpha-\tau_k) \ U(\alpha, \tau_k) \dots U(\beta, \tau_1).$$ 
By definition, $U_{\beta} = \Id$.  We now have the independence in selecting $\tau_1-\tau_k$, in such a way that the corresponding rotation on the $\alpha$ manifold, given by
\begin{equation}
\label{eq:refocus1}
\underbrace{U(\alpha, t_\beta-\tau_1) \dots U(\beta, t_\alpha-\tau_k)}{\ } \underbrace{U(\beta, \tau_k) \dots U(\alpha, \tau_1)}{\ } = U_F . 
\end{equation}We rewrite this relation as
\begin{equation}
\label{eq:refocus2}
\Id = \underbrace{U(\beta, \tau_k - t_\alpha ) \dots U(\alpha, \tau_1 - t_\beta)}_{} \ U_F \underbrace{U(\alpha, -\tau_1) \dots U(\beta, -\tau_k)}_{},
\end{equation}
Equation (\ref{eq:refocus2}), can then be written as, 
\begin{equation}
\label{eq:refocus3}
\Id = \underbrace{B U(\beta, \tau_k) \dots \underbrace{A U(\alpha, \tau_1) \ U_F \ U(\alpha, -\tau_1)}_{U_F^1} \dots \dots U(\beta, -\tau_k)}_{U_F^k},
\end{equation} where $A = U(\alpha, -t_\beta)$ and $B = U(\beta, -t_\alpha)$.

Equation (\ref{eq:refocus3}) can be used to determine  $\tau_1 -\tau_k$. 
In equation (\ref{eq:refocus3}), the times $\tau_1 -\tau_k$, are chosen to 
increase the trace of $U_F^k$ to $2$ (the only matrix in $SU(2)$ with trace $2$ is $\Id$). 
These can be obtained iteratively in a closed form by the following 
optimization procedure. Suppose $\tau_2 - \tau_k$ is fixed. Now
the Eq. (\ref{eq:refocus3}), can be written as
$$ Tr (U_F^k) = Tr(C \ U(\alpha, \tau_1)\ U_F\ U(\alpha, -\tau_1)), $$ where the 
time $\tau_1$ is chosen to maximize $Tr(U_F^k)$. This optimization then takes the form

\begin{equation}
\label{eq:refocus5}
tr (C U(\alpha, \tau_1) U_F U(\alpha, -\tau_1))
= 2 \{ \cos\frac{\theta}{2} \cos\frac{\theta_c}{2} -
\sin \frac{\theta}{2} \sin \frac{\theta_c}{2} n_c \cdot n(\tau_1) \},
\end{equation} where $U_F = \exp(-i \theta n \cdot \sigma)$ and
$C = \exp(-i \theta_c n_c \cdot \sigma)$. 

In the above expression $\tau_1$ generates rotation around axis $d_\alpha$ and 
only rotates $ n_{\perp}$, the part of $n$, that is perpendicular 
to $d_\alpha$. The above expression is maximized when $n_{\perp}$ 
is aligned or anti-aligned (depending on the sign of 
$\sin \frac{\theta}{2} \sin \frac{\theta_C}{2}$) with $(n_c)_{\perp}$, 
the part of $(n_c)_{\perp}$ that is perpendicular to $d_\alpha$. 
We can write an explicit expression for $\tau_1$.
Let $a$ be a unit vector in the direction of 
$\sin \frac{\theta}{2} (n - (n \cdot d_\alpha)d_\alpha)$, and
$b$ be a  unit vector in the direction of 
$\sin \frac{\theta_c}{2} (n_c - (n_c \cdot d_\alpha)d_\alpha)$, then
$\sin(\tau_1) = (b \times a) \cdot d_\alpha$.
Having determined the optimal value $\tau_1$, we can now proceed to 
maximize $\tau_2$ which is now an optimization of a function like
$Tr(C_1 \ U(\beta, \tau_2)\ U_F^1 \ U(\beta, -\tau_2)), $
and so on. Having found a set of values $\tau_1-\tau_k$, this way, we iterate again.
Each iteration increases the value of 
$tr(U_F^k)$, and therefore this value stabilizes to a limit. If this 
value is not $2$, we can increase $k$ and repeat the iteration.
The resulting set of evolution times $\tau_1-\tau_k$ synthesize a selective rotation 
on the $\alpha$ manifold as in Eq. (\ref{eq:refocus1}). Similarly we can synthesize
a selective rotation on the $\beta$ manifold and therefore any rotation of the kind
given in Eq. (\ref{eq:arbitrary}). Alternatively, we can synthesize a desired unitary 
transformation $U_\beta$ on the beta manifold as described in Eq. 
(\ref{eq:rot.tracing}), which results in some rotation $U_\alpha'$ on the alpha 
manifold (that is easily computed).
Now performing an $\alpha$ manifold selective rotation of $U_\alpha (U_\alpha')^{-1}$,
will result in the the desired rotations on the two manifolds. Using the above described method, 
we explicitly compute the pulse sequence for implementing a controlled rotation 
$\exp(-i \pi S^\alpha I_z)$, which inverts the nuclear spins in the 
$\alpha$ manifold and leaves the $\beta$ manifold unaffected. We assume 
$\frac{\omega_I}{A}=.5$ and $\frac{B}{A} = 1$.  This gives 
$t_\alpha = 4 \pi$ and $t_\beta = \frac{4 \pi}{\sqrt{5}}$. The pulse sequence 
$$ \underbrace{\tau_1-\pi-\tau_2 \hdots \tau_5-\pi-t_\alpha-\pi-(t_\beta-\tau_5)-\pi-
(t_\alpha-\tau_4)-\hdots-\pi-(t_\beta-\tau_1)}_{P}, $$ where the switching times, in 
the units of $1/A$ are $\tau_1 = 10.32$, $\tau_2 = 4.64$, $\tau_3=9.75$, $\tau_4=.2760$, 
$\tau_5=10.97$ and $\tau_6=5.47$, implements the desired rotation. Also note that 
using the above sequence as a building block, the pulse sequence $P-\pi-P-\pi$ implements 
the rotation $\exp(-i \pi I_z)$. Similarly we can find a switching sequence $Q$, that implements
$\exp(-i \pi I_x)$. Then the pulse sequence $P-Q-P-Q$ will implement the rotation
$\exp(-i \pi 2 I_zS_z)$.


\begin{remark}({\bf Synthesizing Arbitrary Unitary Transformations:})
{\rm Any unitary transformation on the electron-nuclear spin system is given by \cite{KBG:2001}

$$U_I U_S  \exp(-i[a S_xI_x + bS_yI_y + cS_zI_z]) V_I V_S $$

where $U_I , V_I $ are single spin operations on the  nucleus and $ U_S , V_S $
are single spin operations on the electron. The operators, 
$I_xS_x, I_yS_y, I_zS_z$ all commute. 

As described in Eq. (\ref{eq:arbitrary}), we have shown that all 
rotations of the kind $\exp(-i \theta_1 S_z I_a)$ and $\exp(-i  \theta_2 I_b)$ can be 
synthesized by simply toggling electron spins (here $I_a$ and $I_b$ are 
arbitrary nuclear spin operators). These rotations, along with single spin
rotations, $\exp(-i \theta_3 S_{\gamma})$ on the electron can be used to perform arbitrary 
unitary rotation on the electron-nuclear spin system. The bilinear 
rotations $\exp(-i \theta S_x I_x)$ can be synthesized as 
$$ \exp(-i \theta S_x I_x) = \exp(-i \frac{\pi}{2} S_y) \exp(-i \theta S_zI_x) 
\exp(i \frac{\pi}{2} S_y). $$ }
\end{remark}

\begin{remark}({\bf Refocusing Resonance Offsets for Electron Spins:})
{\rm In the above described methods, we have assumed that 
the precession frequency of the electron is precisely defined and 
there are no resonance offsets. The dispersion in the frequency 
of the electron can be refocused by the following techniques.
For $k$ even, let $\tau_1-\pi-\tau_2-\dots-\tau_k-\pi$, be a 
sequence that implements the propagator 
$ U_f = \exp(-i \frac{b}{2} I_q ) $. Then the sequence $P-P$, 
where $P = \tau_1-\pi-\tau_2-\dots-\tau_k$, will implement the
propogator $\exp(-i b I_q)$ and refocus all resonance effects.

We now consider the problem of synthesizing a general rotation of the kind 
$\exp(-i (a S_zI_p))$, in the presence of resonance offsets.
Let $I_q$ be orthogonal to $I_p$. Let us consider
a switching sequence that implements  $\exp(-i \pi I_q)$ compensating for offset effects
and a switching sequence (assuming no resonance offset),
$Q = \tau_1-\pi-\tau_2-\dots-\tau_k$, such that $k$ is even 
and $Q-\pi$ implements $\exp(-i \frac{a}{2} S_zI_p ) $. Then
the sequence $Q-R-Q-R$ will implement the desired propagator
$\exp(-i a S_z I_p )$ and compensate for offset effects on the electron.}
\end{remark}

\begin{remark}({\bf Polarization Transfer from Electron to Nucleus:})
{ \rm The above described methods can now be used to transfer polarization from 
electrons to the nucleus.  The desired transformation 
$$ S_z \rightarrow I_z, $$ can be performed as follows. 
\begin{equation}
S_z \xrightarrow {\exp(-i\frac{\pi}{2}S_y)} S_x  \xrightarrow {\exp(-i\frac{\pi}{2}2S_zI_y)} 2S_yI_y  \xrightarrow{\exp(-i\frac{\pi}{2}S_x)} 2S_zI_z \xrightarrow{\exp(-i\frac{\pi}{2}2S_zI_x)} I_z , \end{equation}where the various unitary transformations can be performed in a way that are robust to resonance offsets. Furthermore all Bilinear Hamiltonians above can be synthesized by simply doing $\pi$ pulses on the electron as described before. }
\end{remark} 

\begin{remark}({\bf Electron Coupled to Many Nuclei:}){\rm \ We now consider the coupling 
Hamiltonian of a single electron $\frac{1}{2}$, coupled to many spin $\frac{1}{2}$, nuclei. 
The Hamiltonian for the spin system takes the form \cite{SJ:2001}
$$ H_0 = \Omega_S S_z + \sum_{j=1}^N \omega_{jI}I_{jz} + S_z \sum_{j=1}^N (A_j I_{jz} + B_{j}I_{jx}).$$

For now, we neglect the resonance offset $\Omega_S$, for the electron.
As before, for each electron-nuclear coupling, we can define the unit directions 
$(d_{j \alpha}, d_{j \beta})$ and the precession frequencies 
$(\omega_{j \alpha}, \omega_{j \beta})$. Then a straightforward computation 
shows that if for all $j$, the 
precession frequencies $\omega_{j \alpha} \neq \omega_{j \beta}$ and 
the frequency pairs $(\omega_{j \alpha}, \omega_{j \beta})\neq (\omega_{k \alpha}, \omega_{k \beta})$ are distinct, then all the commutators generated by $-iH_1$ and $-iH_2$, where
$$ H_1 = \sum_{j=1}^N \omega_{jI}I_{jz} + S_z \sum_{j=1}^N (A_j I_{jz} + B_{j}I_{jx});\ 
H_2 = \sum_{j=1}^N \omega_{jI}I_{jz} - S_z \sum_{j=1}^N (A_j I_{jz} + B_{j}I_{jx}), $$
span the space $\{I_{j p}, S_z I_{j q} \}$. Therefore any rotation on the 
individual nuclei is achievable by simply switching the electron state by $\pi$ pulses. 
These rotations 
combined with rotations on the electron are sufficient to synthesize any unitary 
transformation on the coupled electron-nuclear spin system. Let $U_{\alpha j}$ and
 $U_{\beta j}$ represent the desired unitary transformation on the $\alpha$
and $\beta$ manifold of the $j^{th}$ nuclei. Let $\tau_1-\pi-\dots-\pi-\tau_k-\pi$, ($k$ even) be
the pulse sequence that implements such a transformation. Then the switching times can 
be computed by iterative maximization over $\tau_1-\tau_k$, of a trace function like
$$ \sum_j tr(U_{\alpha j}'U(\beta_j, \tau_k)\dots U(\alpha_j, \tau_1)) +  tr(U_{\beta j}'U(\alpha_j, \tau_k)\dots U(\beta_j, \tau_1)), $$ where  $U(\alpha_j, \tau)$ represents rotation of 
the $j^{th}$ nuclei around the axis $\alpha_j$, with angular frequency $\omega_{\alpha j}$
for time $\tau$. Gradient algorithms for maximization of such trace functions were presented in 
\cite{ng:grape}}\end{remark}

\section{Conclusion}
In this article, we developed new techniques for 
control of coupled electron-nuclear spin system. 
We showed that the spin system at magnetic field strengths where the Larmor frequency 
of the nucleus is comparable to the hyperfine coupling strength can be controlled by 
simply performing a sequence of flips (each followed by a suitable delay) on the state 
of the electron spin. We showed that this mode of switched control, allows for universal 
control of the nuclear spin states. This switched control of nuclear spin states is much faster
in settings where the Hyperfine coupling is much larger than the Rabi frequency of the nucleus.
Combined with single spin rotations 
of the electron spin, the switched mode control can be used to synthesize any 
unitary transformation on the coupled spin system. We presented algorithms for explicitly
computing the switching times. Furthermore, we discussed how these methods can be made 
robust to  resonance offsets of the electron spin. Further method development is 
required to make these methods robust to dispersion in strength of hyperfine couplings.
Application of these methods for transferring polarization from electron to nuclear spins 
was also discussed. The 
switched control methods presented in this work are expected to find applications 
in pulsed EPR experiments and quantum information processing systems based on 
coupled electron-nuclear spin dynamics.

\begin{acknowledgments}
This work was supported by ONR 38A-1077404, AFOSR FA9550-05-1-0443, and 
NSF 0133673.
\end{acknowledgments}



\bibliography{electrospin}

\end{document}